\documentclass[12pt]{iopart}
\usepackage{iopams}
\usepackage{graphicx}
\usepackage{bm}
\usepackage{subfig}
\usepackage{tabularx}
\usepackage{cases}
\usepackage{ulem}
\makeatletter

\newcommand{\Rmnum}[1]{\expandafter\@slowromancap\romannumeral #1@}
\makeatother

\begin{document}
\title[]{Full quantum treatment of Rabi oscillation driven
by a pulse train and its application in ion-trap
quantum computation}

\author{Li Yang$^1$, Biyao Yang$^1$ and Yufu Chen$^2$}

\address{$^1$ State Key Laboratory of Information Security, Institute of Information Engineering, Chinese Academy of Sciences, Beijing 100093, China}
\address{$^2$ College of Mathematical Science, University of Chinese Academy of Sciences, Beijing 100049, China}
\ead{yangli@iie.ac.cn}

\begin{abstract}
Rabi oscillation of a two-level system driven by a pulse
train is a basic process involved in quantum computation. We
present a full quantum treatment of this process and show that
the population inversion of this process collapses exponentially,
has no revival phenomenon, and has a dual-pulse structure in
every period. As an application, we investigate the properties of
this process in ion-trap quantum computation. We find that in the
Cirac--Zoller computation scheme, when the wavelength of the
driving field is of the order $10^{-6}$ m, the lower bound of failure
probability is of the order $10^{-2}$ after about $10^2$ controlled-NOT gates. This
value is approximately equal to the generally-accepted threshold in fault-tolerant
quantum computation.

\end{abstract}



\maketitle

\section{Introduction}
The quantum algorithms presented show that quantum computation (QC) can solve several problems that are notoriously intractable on classical computers \cite{Shor1994}, and challenge most public-key cryptosystems in use \cite{RSA1978,ElGamal1985}. Many proposals for implementing QC have been put forward. Among them, the cold ion-trap scheme (Cirac--Zoller scheme) \cite{Cirac&Zoller1995} is the earliest and most promising, e.g., a scalable, multiplexed ion trap for quantum information processing has been demonstrated \cite{Leibrandt2009}. Implementation of quantum logic gates in this scheme is realized via Rabi oscillation of ions driven by a pulse train of laser fields. The interaction of a single atom with a radiation field is a basic interaction in physics. In \cite{feng89}, a nonperturbative, fully quantum-theoretical analysis describing the transient spontaneous emission of an initially excited two-level atom in a one-dimensional cavity with output coupling is presented. In \cite{Schadwinkel00}, observations of the quantum dynamics of an isolated neutral atom stored in a magneto-optical trap are presented.

The theoretic measure in \cite{Cirac&Zoller1995} is a typical one
that considers the laser field as a classical field. However,
considering the quantum nature of the driving field, one may obtain
results that differ from those derived through classical
treatment. There are generally two ways to take the
quantum nature of a field into consideration. One is to add quantum fluctuations
to the classical treatment \cite{Julio2002}. However, there are many operations in QC, and the suitability of this method for many operations is not yet determined. The
other way is to quantize the field and calculate the
result\cite{Yang2007}. To do this, we should first consider the Rabi
oscillation driven by a quantized pulse train. This is a basic
atom--photon interaction process, and QC is
one of its many applications. We can then analyze and discuss the failure
probability in ion-trap QC.

Rabi oscillation driven by a quantized continuous-wave (cw)
field, accompanied by collapse-revival phenomenon
\cite{Cummings1965, Eberly1980,Narozhny1981,Knight1982}, is a
typical phenomenon of atom--photon systems. However, Rabi
oscillation driven by a quantized pulse train has not been fully
investigated. It may have different phenomena
from those driven by a cw field.

Fault-tolerant quantum computation (FTQC) allows the computer
to work normally, even when its elementary components are
imperfect. However, the threshold theorem in FTQC requires the
failure probability of each component to be below some threshold
\cite{Nilsen&Chuang2000}. We can then compare the failure
probability of QC with the threshold value, and reach some
meaningful conclusion\cite{Yang2007}.

This paper is arranged as follows: in Section \ref{a1} we
describe a method to deal with the quantum transformation of a
two-level system after one coherent pulse, which expresses the
relationship between the density matrices for the two-level
system before and after one coherent pulse. In Section \ref{a2}
we investigate the properties of Rabi oscillation driven by a
pulse train. In Section \ref{a3} we describe this kind of Rabi
oscillation in ion-trap QC and obtain the
failure probability. In Section \ref{a4} we offer some
discussion, and some conclusions are presented in Section \ref{a5}.

\section{Quantum transformation of a two-level system involving one coherent pulse}\label{a1}

\subsection{Modeling}

The two-level system driven by repeated pulses is an open system, and the usual way to deal with such a system is by Kraus summation and the master equation method. However, for the specific problem here, which cannot be easily solved with those methods, we use the following method: after a single pulse, we obtain the density matrix for the whole system (including a two-level system and the laser field), then obtain the reduced density matrix for the two-level system. We can obtain the relation for the state of the two-level system before and after the pulse, and then the state of the two-level system after repeated pulses can be obtained.

In \cite{Julio2002,Enk&Kimble2002}, the
Jaynes-Cummings model (JCM) \cite{Jaynes&Cummings1963} is used for the
case in which an atom in free space interacts with a laser
field. However, the JCM is a model for describing the
interaction of an atom and a single-mode field in a cavity.
Actually, there is some discussion \cite{Itano2003,Julio2003,
Enk2003} on the validation of the JCM in the multi-mode case.
For example, in a paper by Enk and Kimble
\cite{Enk&Kimble2002}, in Section 2.3 ``Atom-light interaction'',
the case in which an atom in free space interacts
with a laser field is considered, making use of the Hamiltonian of the JCM in
Eq.(10). Enk and Kimble also point out that the Hamiltonian in Eq. (10)
in their paper is valid for atoms in free space for less than one
Rabi period, although a strict proof is not provided.

We analyze the situation as follows: the sources of decoherence can
generally lead to a certain failure probability on a
single qubit or a pair of qubits. After many operations on the
same qubit (or the same pair of qubits), the failure
probability will generally accumulate to reach the threshold in
the threshold theorem of FTQC. The corresponding operation
number is the upper bound of the operation number in one
error-correction period when the given source of decoherence
exists. For many sources of decoherence, such as fluctuation of laser intensity and frequency, beam pointing instabilities, and fluctuation of a magnetic field, the upper bound can be increased by improving the technique. For example, for laser frequency fluctuation, when better frequency stability is achieved, the upper bound for the operation number can be increased to a large value, e.g. $10^{30}$, and this large bound generally has little substantial effect on FTQC.

The decoherence caused by field quantization can also provide an
upper bound for the operation number. Unlike the imperfect control mentioned above, which can be improved experimentally, laser field quantization is based directly on fundamental physical laws, and the corresponding upper bound for the operation number cannot be increased by technique improvement. The
calculation of this decoherence should include the interaction
of all modes in the radiation field with the two-level system.
When using the JCM, only one mode of the field is considered, and this
can also give an upper bound $\beta_1$ for the operation number. The
accurate upper bound for the operation number from field
quantization $\beta<\beta_1$, because the spontaneous emission
induced by vacuum modes is not considered in the JCM. Then if we
use the JCM to estimate the upper bound of operation number in one
error-correction period from field quantization, we can obtain
meaningful results. The two-level system driven by pulse train
can be described as
\begin{equation}
H=\hbar g \left(\rme^{{\rmi}\phi}\sigma_+a+\rme^{-{\rmi}\phi}a^\dag\sigma_-\right),
\end{equation}
where $g$ is the coupling constant, $\phi$ is the beam phase, $\sigma_+$ and $\sigma_-$ are the raising and lowering operators of the two-level system, and $a^\dag$ and $a$ the creation and annihilation operators of photons, respectively. Then the unitary time-evolution operation is given by
\begin{eqnarray} \label{e1}
\fl
U (t)=\cos \left(gt\sqrt{a^\dag a+1}\right)|1\rangle\langle1|+\cos \left(gt\sqrt{a^\dag a}\right)|0\rangle\langle0|\nonumber\\
-{\rmi}\left[\rme^{{\rmi}\phi}\frac{\sin \left(gt\sqrt{a^\dag a+1}\right)}{\sqrt{a^\dag a+1}}a|1\rangle\langle0|
+\rme^{-{\rmi}\phi}a^\dag\frac{\sin\left (gt\sqrt{a^\dag a+1}\right)}{\sqrt{a^\dag a+1}}|0\rangle\langle1|\right],
\end{eqnarray}
with $|0\rangle$ and $|1\rangle$ the ground and excited state of the two-level system respectively.

Generally, the initial state of the whole system is $|\psi
\left(0\right)\rangle=\sum _{n=0}^{\infty
}c_{{n}}|n\rangle\otimes\left (
\alpha|0\rangle+\beta|1\rangle\right)$, where
$|c_n|^2=\frac{\rme^{-\bar{n}}\bar{n}^n}{n!}$, and
$|\alpha|^2+|\beta|^2=1$. A single qubit gate is  usually
implemented through a $k\pi$ pulse in Cirac-Zoller scheme,
whose duration $t_0$ satisfies
$gt_0\sqrt{\bar{n}}=\frac{k\pi}{2}$ \cite{Enk&Kimble2002}, with
$\bar{n}$ the mean number of photons in the pulse. After a
$k\pi$ pulse, the state for the two-level system and laser
field is
\begin{eqnarray}\label{b8}
\fl
|\psi_1\rangle=\alpha\left\{\sum _{n=0}^{\infty }c_{{n}}\left[\cos (\frac{k\pi\sqrt{n}}{2\sqrt{\bar{n}}} )|0,n\rangle-{\rmi}\rme^{{\rmi}\phi}\sin (\frac{k\pi\sqrt{n}}{2\sqrt{\bar{n}}})|1,n-1\rangle\right]\right\}\nonumber\\
+\beta\left\{\sum _{n=0}^{\infty }c_{{n}}\left[\cos (\frac{k\pi\sqrt{n+1}}{2\sqrt{\bar{n}}})|1,n\rangle-{\rmi}\rme^{-{\rmi}\phi}\sin (\frac{k\pi\sqrt{n+1}}{2\sqrt{\bar{n}}})|0,n+1\rangle\right]\right\}.
\end{eqnarray}

The corresponding density matrix for the state in (\ref{b8}) is ${ \rho}_{total}^{(1)}=|\psi _1\rangle\langle\psi _1|$. This matrix contains the information for both the two-level system and the field, but we are interested only in the two-level system. Thus we obtain the reduced density matrix $\rho^{ (1)}$, with
\begin{eqnarray*} \label{e3}
{\rho}^{ (1)}_{11}= |\alpha|^2S_4+\frac{{\rm i}}{2} (\alpha\beta^*-\alpha^*\beta){\rm e}^{{\rm i}\phi}S_2+|\beta|^2 (1-S_6),\\
{\rho}^{ (1)}_{12}=\alpha\beta^*S_5+{\rm i} (|\alpha|^2{\rm e}^{{\rm i}\phi}S_1-|\beta|^2{\rm e}^{-{\rm i}\phi}S_7)+\alpha^*\beta S_3,\\
{\rho}^{ (1)}_{21}=\alpha^*\beta S_5-{\rm i} (|\alpha|^2{\rm e}^{{\rm i}\phi}S_1-|\beta|^2{\rm e}^{-{\rm i}\phi}S_7)+\alpha\beta^*S_3,\\
{\rho}^{ (1)}_{22}=|\alpha|^2 (1-S_4)-\frac{{\rm i}}{2} (\alpha\beta^*-\alpha^*\beta){\rm e}^{{\rm i}\phi}S_2+|\beta|^2S_6,\\
\end{eqnarray*}
here
\begin{eqnarray}\label{b4}
&S_1=\sum _{n=0}^{\infty }\frac{{{\rm e}^{-\bar{n}}}\bar{n}^n}{n!}\sqrt{\frac{\bar{n}}{n+1}}\cos (\frac{k\pi\sqrt{n}}
{2\sqrt{\bar{n}}})\sin (\frac{k\pi\sqrt{n+1}}{2\sqrt{\bar{n}}}),\nonumber\\
&S_2=\sum _{n=0}^{\infty }\frac{{{\rm e}^{-\bar{n}}}\bar{n}^n}{n!}\sqrt{\frac{k\bar{n}}{2(n+1)}}
\sin (\frac{k\pi\sqrt{n+1}}{\sqrt{\bar{n}}}),\nonumber\\
&S_3=\sum _{n=0}^{\infty }\frac{{{\rm e}^{-\bar{n}}}\bar{n}^n}{n!}\sqrt{\frac{n}{n+1}}\sin (\frac{k\pi\sqrt{n}}
{2\sqrt{\bar{n}}})\sin (\frac{k\pi\sqrt{n+1}}{2\sqrt{\bar{n}}}),\nonumber\\
&S_4=\sum _{n=0}^{\infty }\frac{{{\rm e}^{-\bar{n}}}\bar{n}^n}{n!}\cos^2 (\frac{k\pi\sqrt{n}}{2\sqrt{\bar{n}}}),\\
&S_5=\sum _{n=0}^{\infty }\frac{{{\rm e}^{-\bar{n}}}\bar{n}^n}{n!}\cos (\frac{k\pi\sqrt{n}}{2\sqrt{\bar{n}}})
\cos (\frac{k\pi\sqrt{n+1}}{2\sqrt{\bar{n}}}),\nonumber\\
&S_6=\sum _{n=0}^{\infty }\frac{{{\rm e}^{-\bar{n}}}\bar{n}^n}{n!}\cos^2 (\frac{k\pi\sqrt{n+1}}{2\sqrt{\bar{n}}}),\nonumber\\
&S_7=\sum _{n=0}^{\infty }\frac{{{\rm e}^{-\bar{n}}}\bar{n}^n}{n!}\sqrt{\frac{n}{\bar{n}}}\cos (\frac{k\pi\sqrt{n+1}}{2\sqrt{\bar{n}}})\sin (\frac{k\pi\sqrt{n}}{2\sqrt{\bar{n}}})\nonumber.
\end{eqnarray}

\subsection{Transforms of the density matrix after a coherent pulse}\label{b2}

Consider the relationship between ${\rho}^{(1)}$ and the density matrix of corresponding initial state
${\rho}^{(0)}=|\psi \left(0\right)\rangle\langle\psi \left(0\right)|.$
For a two-level system, the density matrix ${\rho}$ satisfies the condition ${\rho}=\frac{1}{2} ({I}+{r}\cdot{\sigma})$ \cite{Nilsen&Chuang2000}, ${r}$ is the Bloch vector for state ${\rho}$, $|{r}|$$\leq 1$, ${\sigma}=\left[\begin{array}{lll}\sigma_{\rm x}& \sigma_{\rm y}& \sigma_{\rm z}\\\end{array}\right]^T$.

Let {\footnotesize${r}^{ (m)}=\left[\begin{array}{lll} r^{
(m)}_{\rm x}&r^{ (m)}_{\rm y}&r^{ (m)}_{\rm
z}\\\end{array}\right]^T$} denotes the Bloch vector of ${\rho^{
(m)}}$. An arbitrary trace-preserving quantum operation is
equivalent to a map of the form
${r}\stackrel{\cal{E}}{\rightarrow}{r'}={Mr}+{c}$~\cite{Nilsen&Chuang2000},
here ${M}$ and ${c}$ contain the properties of the system and are
independent of the state. Based on this, it can be seen that
${r}^{ (1)}={Mr}^{ (0)}+{c}$, here ${c}=
\left[\begin{array}{lll}0 & S_7{\rm e}^{-{\rm i}\phi}-S_1{\rm
e}^{{\rm i}\phi}&S_4+S_6-1
\\\end{array}\right]^T$,
\begin{equation*}
{M}=\left[\begin{array}{cc}
S_3+S_5 & 0\\
0 &{M}_1 \\
\end{array}\right],\\
{M_1}=\left[\begin{array}{cc}
S_5-S_3&- ({\rm e}^{{\rm i}\phi}S_1+{\rm e}^{-{\rm i}\phi}S_7) \\
S_2{\rm e}^{{\rm i}\phi}&S_4-S_6\\
\end{array}\right],
\end{equation*}
then ${r}^{(m)}={Mr}^{(m-1)}+{c}$.

\subsection{Calculation of the sums in the density matrix} \label{b1}

It is necessary to get accurate values of $S_i~(i=1,\cdots,7)$
to evaluate the behavior of pulse train. The usual algorithm
(saddle-point approximation) can only reach a precision of
$1/\sqrt{\bar{n}}$. Our algorithm achieving any given precision
instead of the usual algorithm is as follows.

Suppose $\bar{n}$ is not small, for the sum
\[S_i=\sum _{n=0}^{\infty }\frac{{{\rm e}^{-\bar{n}}}\bar{n}^n}{n!}f_{i0} (n,\bar{n},k)
\]

(1) Substitute $n$ in $f_{i0} (n,\bar{n},k)$ with $ (x+1)\bar{n}$ , we get
$f_{i1} (x,\bar{n},k)=f_{i0} \Big((x+1)\bar{n},\bar{n},k\Big).$

(2) Do the Taylor expansion to $x^p$ for $f_{i1} (x,\bar{n},k)$ at $x=0$, and get $f_{i2} (x,\bar{n},k)$.

(3) Since sum $ \sum _{n=0}^{\infty }\frac{{{\rm e}^{-\bar{n}}}\bar{n}^n}{n!}n^k$ can be obtained accurately, we replace $x$ in $f_{i2} (x,\bar{n},k)$ by $\frac{n-\bar{n}}{\bar{n}}$ and get $f_{i3} (n, \bar{n},k)$.

(4) Use $f_{i3} (n, \bar{n},k)$ instead of $f_{i0} (n,\bar{n},k)$ in the expression of $S_{i} (\bar{n},k)$ to calculate the new sum and get $f_{i4} (\bar{n},k)$.

(5) Substituting $\bar{n}$ into $f_{i4} (\bar{n},k)$, we obtain a high-precision result of the original sum $S_i (\bar{n},k)$. The value for $S_i~(i=1,\cdots,7)$ in the cases where we expand $f_{i1} (x,\bar{n},k)$ to $x^{10}$ and $x^{15}$ are compared in Table \ref{t1}.

\begin{table}[h]
\caption{\label{t1} Values for $S_i (i=1,2,\cdots,7)$ for
$\bar{n}=10^4$ and $k=2$. Value1 denotes value of the
resulting sums of the algorithm when we expand $f_{i1}
(x,\bar{n},k)$ to $x^{10}$ and Value2 denotes that when we
expand $f_{i1} (x,\bar{n},k)$ to $x^{15}$. Value1 and Value2
are the same to the precision $10^{-23}$.}
\begin{indented}
\lineup
\item[]
{\scriptsize
\begin{tabular}{@{}ccc}
\br
Sum&Value1&Value2\cr
\mr
$S_1$&$0.000\,039\,303\,916\,656\,063\,668\,561\,519\,091$&$0.000\,039\,303\,916\,656\,063\,668\,561\,194\,770$\cr
$S_2$&$0.000\,039\,265\,164\,255\,300\,772\,996\,074\,590$&$0.000\,039\,265\,164\,255\,300\,772\,995\,750\,283$\cr
$S_3$&$0.000\,246\,659\,192\,761\,352\,167\,541\,307\,293$&$0.000\,246\,659\,192\,761\,352\,167\,542\,402\,758$\cr
$S_4$&$0.999\,753\,309\,972\,685\,637\,856\,777\,333\,369$&$0.999\,753\,309\,972\,685\,637\,856\,776\,237\,858$\cr
$S_5$&$0.999\,753\,316\,133\,881\,571\,308\,212\,070\,145$&$0.999\,753\,316\,133\,881\,571\,308\,210\,974\,684$\cr
$S_6$&$0.999\,753\,322\,301\,165\,250\,291\,025\,614\,276$&$0.999\,753\,322\,301\,165\,250\,291\,024\,518\,866$\cr
$S_7$&$0.000\,039\,226\,416\,698\,193\,975\,826\,600\,887$&$0.000\,039\,226\,416\,698\,193\,975\,830\,095\,264$\cr
\br
\end{tabular}
}
\end{indented}
\end{table}

The precision of the sums $S_i (i=1,2,\cdots,7)$ is ensured by
the following theorem:

{\bf Theorem 1:} For every given integer $l<<\bar{n}$, let
\begin{equation} \label{e2}
p\geq\left\lceil \frac{2\ln\Big(\sqrt{2}\bar{n}^{l+\frac{1}{2}}\sqrt{(l+1)\ln\bar{n}}\Big)}{\ln\bar{n}-\ln[2(l+1)\ln\bar{n}]}
\right \rceil,
\end{equation}
\[
\alpha_0=\frac{1}{\sqrt{\bar{n}}}+\frac{(l+1)\ln\bar{n}}{\sqrt{\bar{n}}}
+\sqrt{\frac{(l+1)^2(\ln\bar{n})^2}{\bar{n}}+2(l+1)\ln\bar{n}}.
\]
If $\alpha_0<\alpha<<\sqrt{\bar{n}}$,
then
\begin{equation} \label{e25}
S_i=\sum_{n=0}^{\infty}\frac{e^{-\bar{n}}\bar{n}^n}{n!}f_{i3}(n,
\bar{n}, k)+o\Big(\frac{1}{\bar{n}^l}\Big),
\end{equation}
here $p$, $f_{i3} (n,\bar{n},k)$ are parameters defined in steps (2) and (3) of the algorithm above. Then this algorithm can reach a precision of $(o(1/\bar{n}^l),l\ll \bar{n})$, much higher than that$(1/\sqrt{\bar{n}})$ of the usual algorithm using the saddle-point approximation \cite{Mandel95,Born99}. For example, when $\bar{n}=10^4$, the usual algorithm can only reach a precision of $10^{-2}$, but for our algorithm, with an appropriate order of Taylor expansion ($p\geq51$), we can easily reach the precision of $10^{-40}$ or higher as needed.

Theorem 1 can be proved using the following three lemmas (see
\ref{a6} for the detailed proof):

{\bf Lemma 1:} For every given $\alpha<<\sqrt{\bar{n}}$,
\begin{equation}  \label{e23}
\sum_{n=\bar{n}-\alpha\sqrt{\bar{n}}}^{\bar{n}+\alpha\sqrt{\bar{n}}}\frac{e^{-\bar{n}}\bar{n}^n}{n!}(f_{i0} (n,\bar{n},k)-f_{i3} (n,\bar{n},k))=o\Big(\frac{\alpha^{p+1}}{(\sqrt{\bar{n}})^{p-1}}\Big).
\end{equation}
here $p$, $f_{i0} (n,\bar{n},k)$, $f_{i3} (n,\bar{n},k)$ are parameters defined in the algorithm above.

{\bf Lemma 2:} For every given integer $l<<\bar{n}$,
$\alpha<<\sqrt{\bar{n}}$,
if $\alpha>\sqrt{(l+1)\ln\bar{n}}$, then
\begin{eqnarray} \label{e4}
\sum_{n=0}^ke^{-\bar{n}}\frac{\bar{n}^n}{n!}<\frac{1}{\bar{n}^l},
\end{eqnarray}
where $k=\left \lceil \bar{n}+\alpha\sqrt{\bar{n}} \right
\rceil.$

{\bf Lemma 3:} For every given integer $l<<\bar{n}$,
$\alpha<<\sqrt{\bar{n}}$, if
\begin{equation}
\alpha>\frac{1}{\sqrt{\bar{n}}}+\frac{(l+1)\ln\bar{n}}{\sqrt{\bar{n}}}+\sqrt{\frac{(l+1)^2(\ln\bar{n})^2}{\bar{n}}+2(l+1)\ln\bar{n}},
\end{equation}
then
\begin{equation}\label{e8}
\sum_{n=k'}^{\infty}e^{-\bar{n}}\frac{\bar{n}^n}{n!}<\frac{1}{\bar{n}^l},
\end{equation}
where $k'=\left \lfloor \bar{n}+\alpha\sqrt{\bar{n}} \right
\rfloor.$

We expand $f_1 (x,\bar{n},k)$ to $x^{15}$ ($p=15$) at $x=0$, and find
the value of the sum is the same to the precision $10^{-23}$ ($l=5$) as
that when we expand $f_1 (x,\bar{n},k)$ to $x^{10}$ ($p=10$). However, the value of $p$ obtained from Eq.~(\ref{e2}) is 24, which implies
that the precision of the sum is much higher than
Eq.~(\ref{e25}) shows. The reason is probably that we have not
considered the periodicity of trigonometric functions, and
the precision of the sum may be considerably improved by the
positive and negative terms canceling each other out.

For small $\bar{n}$, we need only to require $t$ satisfying
\begin{equation} \label{e18}
(t-1)!>e^{-\bar{n}}\bar{n}^{t+l},
\end{equation}
where $t$ is the parameter in sum $S_i (\bar{n},k)=\sum
_{n=0}^{t}\frac{{{\rm e}^{-\bar{n}}}\bar{n}^n}{n!}f_{i0}
(n,\bar{n},k)+o\Big(\frac{1}{\bar{n}^l}\Big)$. For a given
precision $l$, we can search for the smallest $t$ satisfying
(\ref{e18}), e.g., when $\bar{n}=10$ and $l=20$, we get $t=55$.

\section{Population inversion}\label{a2}

\subsection{Final state of the two-level system after pulse train}

Provided ${r}^{(m)}={Mr}^{(m-1)}+{c}$ , then
\begin{eqnarray} \label{e19}
{r}^{ (m)}&={M}^m{r}^{(0)}
+({M}^{m-1}+\cdots+{M}+{I}){c}.
\end{eqnarray}
It can be seen from Sec.~\ref{b2} that
\[{M}^m=\left[\begin{array}{ccc}
(S_3+S_5)^m & 0\\
{O} &{M}_1^m \\
\end{array}\right],{M_1}=\left[\begin{array}{ccc}
S_5-S_3&- (S_1+S_7) \\
S_2&S_4+S_6-1\\
\end{array}\right],\]
For any real matrix ${A}=\left[\begin{array}{ccc}
a&b\\
c&d \\
\end{array}\right],$
we obtain (see \ref{a7})
\begin{equation} \label{e21}
{A}^m=\frac{\Lambda_{+}^{(m)}}{2}{I}+\frac{\Lambda_{-}^{(m)}}{2\rmi Q}\left[\begin{array}{ll}
-K&2b\\
2c&K\\
\end{array}\right],
\end{equation}
where $\Lambda_{\pm}^{(m)}=\lambda_1^m\pm\lambda_2^m$ with
$\lambda_1$ and $\lambda_2$ eigenvalues of $A$, $K=d-a,
Q=-{\rmi}\sqrt{(a-d)^2+4bc}$. When $(a-d)^2+4bc<0$ (which is
the case for $M_1$)
\begin{equation*}
{A}^m=|\lambda|^m\Big[\cos(m\theta){I}+\sin(m\theta)
\frac{{J}}{\sqrt{\det{J}}}\Big],
\end{equation*}
where $|\lambda|^2=ad-bc$, $ \sin\theta=\frac{1}{2}\sqrt{2-\frac{a^2+d^2+2bc}{ad-bc}},$ ${J}=\left[\begin{array}{cc}
a-d&2b\\
2c&d-a\\
\end{array}\right].$
Therefore,
\[{I}+{M}_1+\cdots+{M}_1^{m-1}=\Big[\sum _{j=0}^{m-1 }|\lambda|^j\cos(j\theta)\Big]{I}+\Big[\sum _{j=0}^{m-1 }|\lambda|^j\sin(j\theta)\Big]\frac{{J}}{\sqrt{\det{J}}}.\]
Since
\begin{eqnarray*}
\sum _{j=0}^{m-1 }\{|\lambda|^j[\cos(j\theta)&+\rmi\sin(j\theta)]\}=\sum _{j=0}^{m-1 }(|\lambda|^j\rme^{\rmi(j\theta)})=\sum _{j=0}^{m-1 }(|\lambda|\rme^{\rmi\theta})^j\\
&=\frac{1-|\lambda|\rme^{-\rmi\theta}-|\lambda|^m\rme^{\rmi(m\theta)}
+|\lambda|^{m+1}\rme^{\rmi(m-1)\theta}}{1+|\lambda|^2
-2|\lambda|\cos\theta},
\end{eqnarray*}
we have
\begin{eqnarray*}
\sum _{j=0}^{m-1 }|\lambda|^j\cos(j\theta)&=\frac{1-|\lambda|\cos\theta
-|\lambda|^m\cos(m\theta)
+|\lambda|^{m+1}\cos(m-1)\theta}{1+|\lambda|^2
-2|\lambda|\cos\theta},\\
\sum _{j=0}^{m-1}|\lambda|^j\sin(j\theta)&=\frac{|\lambda|
\sin\theta-|\lambda|^m\sin(m\theta)
+|\lambda|^{m+1}\sin(m-1)\theta}{1+|\lambda|^2
-2|\lambda|\cos\theta},
\end{eqnarray*}
thus
\begin{eqnarray*} \label{e22}
\fl
&{I}+{M}_1+\cdots+{M}_1^{m-1}\\&=\frac{1}{1+|\lambda|^2
-2|\lambda|\cos\theta}\Big[(1-|\lambda|
\cos\theta-|\lambda|^m\cos(m\theta)
+|\lambda|^{m+1}\cos(m-1)\theta){I}\\&+(|\lambda|
\sin\theta-|\lambda|^m\sin(m\theta)
+|\lambda|^{m+1}\sin(m-1)\theta)\frac{{J}}{\sqrt{\det{J}}}\Big]\nonumber\\
&\stackrel{\triangle}{=}B_1^{(m)}{I}+B_2^{(m)}{J},
\end{eqnarray*}
then
\begin{eqnarray*}
{r}^{ (m)}&=\left[\begin{array}{cc}
(S_3+S_5)^m&0 \\
{O}&|\lambda|^{\frac{m}{2}}\Big[\cos(m\theta){I}+\sin(m\theta)
\frac{{J}}{\sqrt{\det{J}}}\Big]\\
\end{array}\right]{r}^{(0)}\nonumber\\
&+\left[\begin{array}{cc}
\frac{1-(S_3+S_5)^m}{1-S_3-S_5}&0 \\
{O}&B_1^{(m)}{I}+B_2^{(m)}{J}\\
\end{array}\right]{c}.
\end{eqnarray*}

\subsection{Population inversion after pulse train}

Suppose the initial state is $|1\rangle$, if we have applied $k\pi$ pulses for m times, the population inversion is
\[
W_m=\frac{1}{2}(1-r_{\rm z}^{(m)})-\frac{1}{2}(1+r_{\rm z}^{(m)})=-r_{\rm z}^{(m)},
\]
we have
\begin{eqnarray}
\fl
W_m=|\lambda|^m\Big[\cos(m\theta)+\sin(m\theta)
\frac{j_{22}}{\sqrt{\det{J}}}\Big]
-\frac{1}{1+|\lambda|^2
-2|\lambda|\cos\theta}\nonumber\\
\times\Big\{\Big[|\lambda|\sin\theta-|\lambda|^m\sin(m\theta)
+|\lambda|^{m+1}\sin(m-1)\theta\Big]\frac{j_{21}}{\sqrt{\det{J}}}(S_7-S_1)\nonumber\\
+\Big[\Big(1-|\lambda|\cos\theta-|\lambda|^m\cos(m\theta)
+|\lambda|^{m+1}\cos(m-1)\theta\Big)\nonumber\\
+\Big(|\lambda|\sin\theta-|\lambda|^m\sin(m\theta)
+|\lambda|^{m+1}\sin(m-1)\theta\Big)\frac{j_{22}}{\sqrt{\det{J}}}\Big](S_4-S_6)\Big\}.
\end{eqnarray}

To obtain the inversion between the $m$th and $(m+1)$th $k\pi$
pulse, we should first obtain the corresponding density matrix of
the two-level system
\begin{equation}
\rho_m(t)=\tr\left\{U(t)[\rho^{(m)}\otimes\rho_{\rm l}]U^{\dag}(t)\right\},
\end{equation}
where $U(t)$ is the unitary time-evolution operator mentioned earlier, and $\rho_{\rm l}$ is the density matrix for the laser field. A detailed calculation gives the probability that the ion is in state $|0\rangle$:
\begin{eqnarray*}
&p(t)=\frac{1}{2}\Big[(S_8+S_9)+r_{\rm z}^{(m)}(t)(S_8-S_9)+r_{\rm y}^{(m)}(t)S_{10}\Big],
\end{eqnarray*}
where
$S_8=\sum_{n=0}^{\infty}\frac{{\rm e}^{-\bar{n}}\bar{n}^n}{n!}\cos^2gt\sqrt{n}$,
$S_9=\sum_{n=0}^{\infty}\frac{{\rm e}^{-\bar{n}}\bar{n}^n}{n!}\sin^2gt\sqrt{n+1}$
and $S_{10}=\sum_{n=0}^{\infty}$ \\$\frac{{\rm e}^{-\bar{n}}\bar{n}^n}{n!}\sqrt{\frac{n}{\bar{n}}} \sin2gt\sqrt{n}$,
All these values can be obtained with high precision using our algorithm in
Section \ref{b1}, and the inversion is
$W_m(t)=1-2p(t),0<t<\frac{k\pi}{2g\sqrt{\bar{n}}}$.

Consider the difference between the oscillation driven by
pulse train and by a cw field. The population inversion for
repeated $2\pi$ pulses is shown in Fig.1 (given
$\bar{n}=10$). We find there is a dual-pulse structure in every
period, where the amplitude starts to increase from the point
of $2\pi$ pulses. The inversion decreases exponentially,
unlike a Gaussian function collapse envelope driven by
a cw field. Besides, there is no revival phenomenon, but a
small nonzero amplitude exists. The reason for this behavior can be analyzed as follows:
when a laser field comes to drive a two-level system, the infinite number state components become entangled with the state of the two-level system, and the state for the two-level system and the laser field can be written as
{\footnotesize \begin{equation*}
|\varphi_1\rangle=\sum_{n=0}^{\infty }\Big(A_1(n)|0\rangle |n\rangle+A_2(n)|1\rangle |n-1\rangle+A_3(n)|1\rangle |n\rangle+A_4(n)|0\rangle |n+1\rangle\Big).
\end{equation*}}
Then the two-level system's state becomes mixed, which can be written as $\sigma^{(1)}=\sigma^{(1)}_{11}|0\rangle\langle0|+\sigma^{(1)}_{12}|0\rangle\langle1|+\sigma^{(1)}_{21}|1\rangle\langle0|+\sigma^{(1)}_{22}|1\rangle\langle1|$, and between each
component there is no fixed relation in phase. Thus, when another laser field comes to interact
with the two-level system, each number state component of the laser independently entangles with each component of the two-level system's state, thus forming a more complicated mixed state
\begin{eqnarray*}
|\varphi_2\rangle&=U(t)\Big(
\sigma^{(1)}_{11}\sum_{n=0}^{\infty }\frac{{\rm e}^{-\bar{n}}\bar{n}^n}{n!}|0\rangle\langle0|\otimes |n\rangle\langle n|+\sigma^{(1)}_{12}\sum_{n=0}^{\infty }\frac{{\rm e}^{-\bar{n}}\bar{n}^n}{n!}|0\rangle\langle1|\otimes |n\rangle\langle n|\nonumber\\
&+\sigma^{(1)}_{21}\sum_{n=0}^{\infty }\frac{{\rm e}^{-\bar{n}}\bar{n}^n}{n!}|1\rangle\langle0|\otimes |n\rangle\langle n|+\sigma^{(1)}_{22}\sum_{n=0}^{\infty }\frac{{\rm e}^{-\bar{n}}\bar{n}^n}{n!}|1\rangle\langle1|\otimes |n\rangle\langle n|\Big)U(t)^\dag,
\end{eqnarray*}
where $U(t)$ is given in Eq.(\ref{e1}).After many iterations of interaction with different laser fields, the final state of the two-level system becomes an extremely complicated mixed state, and has little initial phase information.

From the point of dissipation in a quantum open system, this phenomenon can be understood as follows: the existence of revival in Rabi oscillation
 driven by a cw field is because the asynchronous probability amplitude (which causes collapse) becomes synchronous in phase again after a period of time. This ``memory effect" of the oscillation phase comes about because that the phase information is kept in the driving laser field, and the laser field is still in the cavity. However, for Rabi oscillation driven by a pulse stream that is an open system, after tracing out the environment (laser field), the master system (two-level system) loses its phase information. The physical picture is that a laser field leaves the two-level system and takes away the phase information after interacting with it. Then after many iterations of interaction with different pulses, the phase information is lost repeatedly (dissipation), and finally there is no revival phenomenon.
\begin{figure}[h]
\centering
\subfloat[]{
\begin{minipage}[h]{1\textwidth}
\centering
\label{figc}
\includegraphics[width=5in,height=2.7in]{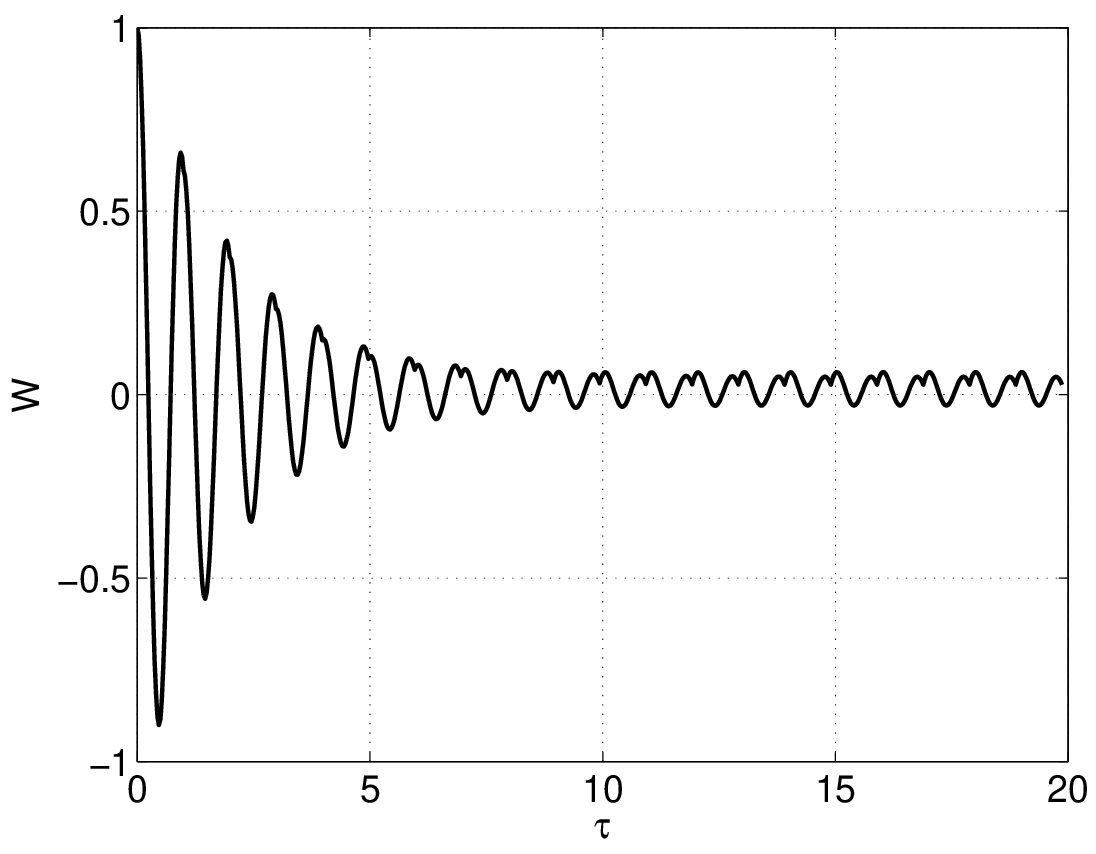}
\end{minipage}
}
\newpage
\subfloat[]{
\begin{minipage}[h]{1\textwidth}
\centering
\label{figd}
\includegraphics[width=5in,height=2.7in]{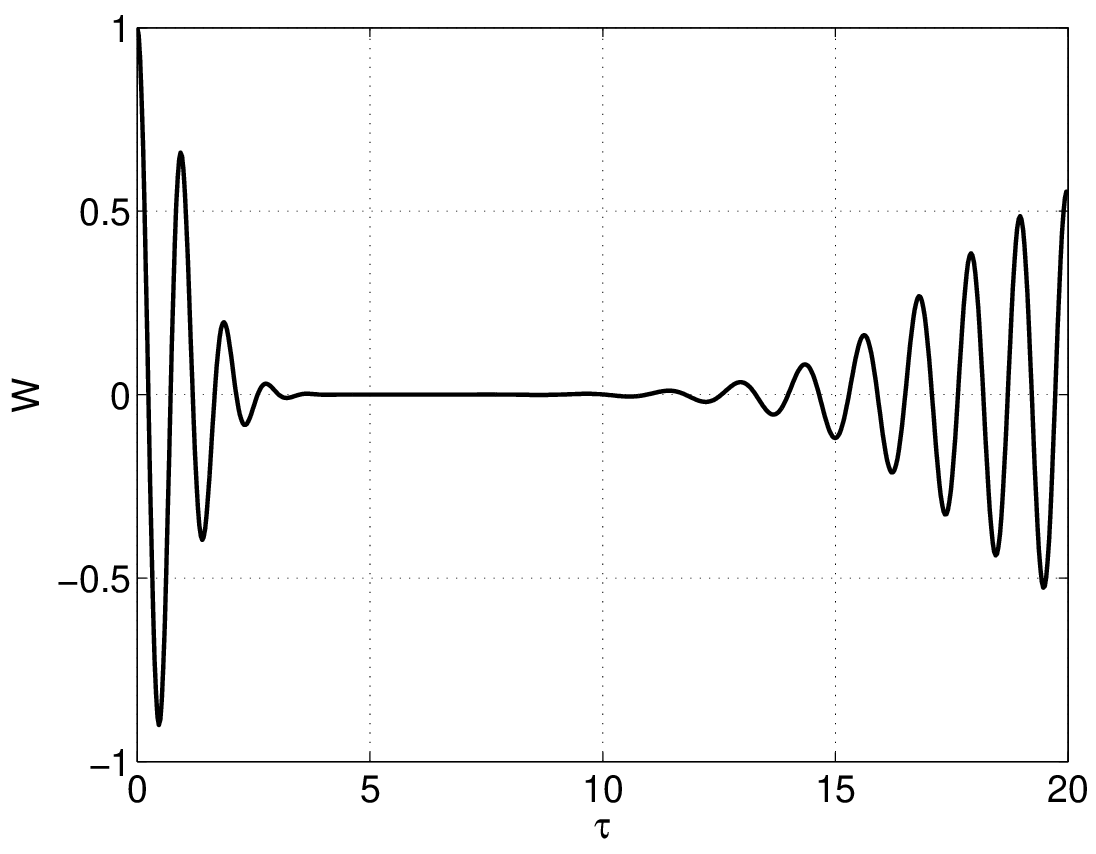}
\end{minipage}
} \caption{Population inversion driven by different fields,
given $\bar{n}=10$, $\tau=gt$. (a) $2\pi$ pulse train case.
There is a dual-pulse structure in every period, where the
amplitude starts to increase from the point of $2\pi$ pulses.
The inversion decreases exponentially, unlike a
Gaussian function collapse envelope driven by a cw field.
Besides, there is no revival phenomenon, but a small nonzero
amplitude exists (the amplitude of each dual-pulse structure's crest approaches to a stable value). (b) Corresponding inversion driven by a cw
field.}
\end{figure}

\begin{figure}[h]
\centering
\includegraphics[width=3in]{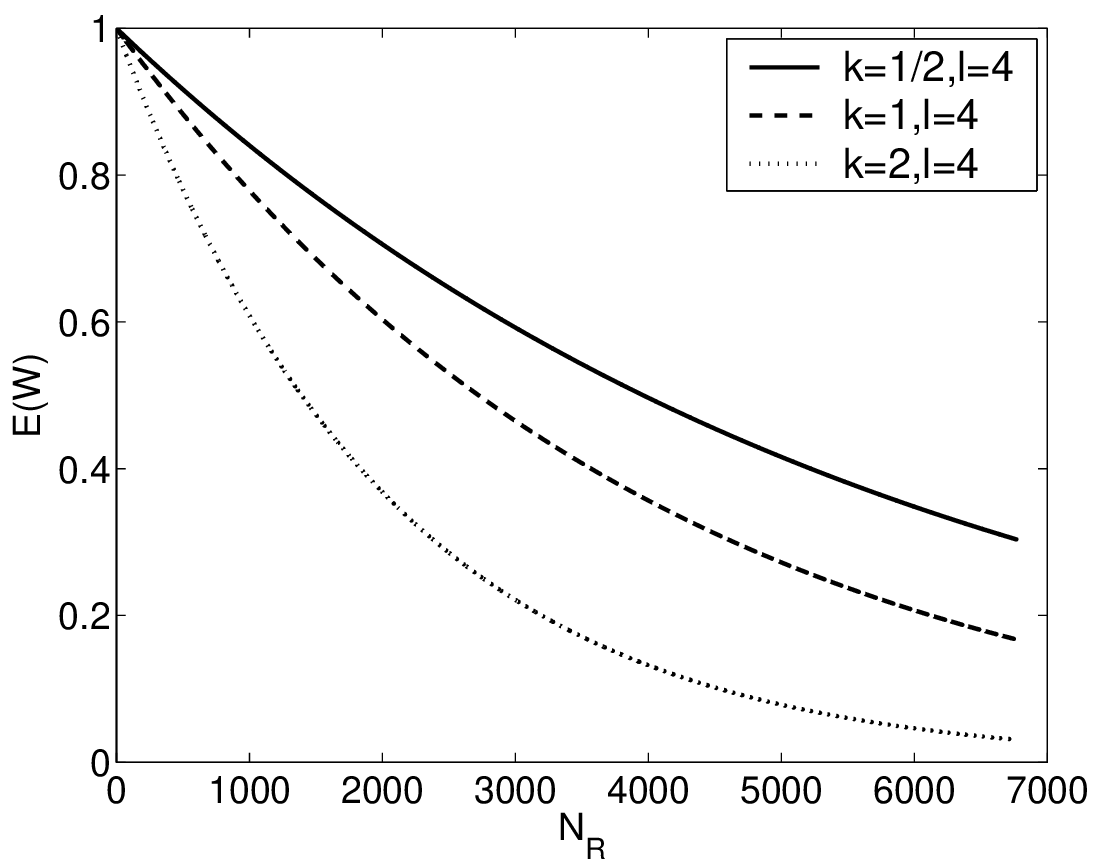}
\caption{\label{fig4}
Inversion at the points of $2\pi$ pulses $E(W)$ versus number of Rabi periods $N_{\rm R}$, $\bar{n}=10^4$. Fitting results are $1.0031\rme^{-0.0002N_{\rm R}}$, $1.0193\rme^{-0.0003N_{\rm R}}$, $1.025\rme^{-0.0005N_{\rm R}}$ for $k=1/2, 1, 2$ respectively.}
\end{figure}

The inversion at the points of $2\pi$ pulses when $\bar{n}=10^4$ is plotted in Figure \ref{fig4}. Results of fitting is $1.0031{\rm e}^{-0.0002N_{\rm R}}$, $1.0193{\rm e}^{-0.0003N_{\rm R}}$, $1.025{\rm e}^{-0.0005N_{\rm R}}$ for $k=1/2, 1, 2$ respectively, here $N_{\rm R}$ is the number of Rabi periods.

\section{Failure probability of gate operation realized through Rabi oscillation driven by repeated pulses}\label{a3}

\subsection{Estimation of $\bar{n}$}

The value $\bar{n}$ that determines $S_i$ is an important parameter in our discussion. To estimate the mean number of photons in one pulse, we assume a fictitious pulse is propagating simultaneously in the opposite direction. They instantly form a standing wave when overlapping in space. It can be seen that the mean number of photons in each pulse is about half of thoes in the standing wave. We now focus on the mean number of photons in the imaginary standing wave.

The electric field $E$ can be expressed as $E=\mathcal{E} \sqrt{\bar{n}}.$
$\cal{E}$ is usually given as $\cal{E}$=$\sqrt{\frac{\hbar\omega}{\epsilon_0V}}$ \cite{Sargent1974}, where $\omega$ is the frequency of the single mode in a cavity, and $V$ is the volume of the cavity. It can be seen that $V\sim Act$, with $A$ the cross-sectional area of the beam, thus
$\bar{n}=\frac{\epsilon_0Act}{\hbar\omega}E^2.$
For a $k\pi$ pulse, $gt\sqrt{\bar{n}}=\frac{k\pi}{2}$,  $g\sim\frac{p\cal{E}}{\hbar}=\frac{pE}{\hbar\sqrt{\bar{n}}}$, with $p\sim {\rm e}a_0$ the electric dipole moment of the ion, e the charge of an electron, and $a_0$ the Bohr radius, then we obtain $t=\frac{k\pi\hbar}{2pE}.$
Then we have $\bar{n}=\frac{k}{4}\frac{\epsilon_0A\lambda}{p}E$.

Any photon in a beam has a probability amplitude at every point of the beam's cross-sectional area. Then, when a laser beam (beam A) interacts with a trapped ion, all the photons interact with the ion. However, only the probability amplitude in an ``effective interaction area" (around the ion) is useful for the interaction. Thus, this in some sense is equivalent to a beam (beam B) with ``effective interaction area" interacting with the ion, where any photon's probability amplitude at every point of the beam is useful for interaction. Then the mean number of photons in beam B is in effect the mean number of photons in beam A. One may take the total resonant scattering cross-section for an atomic dipole transition as the effective interaction area, but when a photon is scattered in the paraxial mode, there is actually no interaction. Then the effective interaction area is the cross section for scattering out of the paraxial modes.

Now we calculate the effective mean number of photons. When a laser beam is applied to a trapped ion, the total resonant scattering cross section for an atomic dipole transition is $\sigma=3\lambda^2/2\pi$ \cite{Cohen-Tannoudji1992}, and the cross section for scattering out of the paraxial modes is $\sigma_{eff}=3\lambda^2/8\pi$ \cite{Silberfarb2003}. Then the effective interaction area is $\sigma_{eff}$, and the photons in volume $\sigma_{eff}ct$ is effective. For each photon, the probability of being in area $\sigma_{eff}$ is $\frac{\sigma_{eff}}{A}$, and the probabilities are independent for the photons. It can be seen that the effective mean number of photons is
\begin{equation}  \label{b16}
\bar{n}_{eff}=\bar{n} \frac{\sigma_{eff}}{A}=\frac{k}{4}\frac{\epsilon_0\sigma_{eff}\lambda}{p}E.
\end{equation}

\begin{figure}[htbp]
\centering
\includegraphics[width=4in]{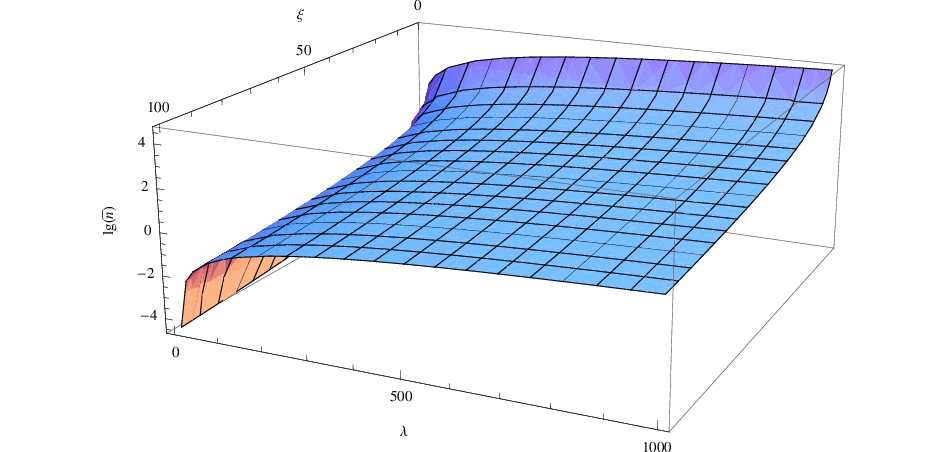}
\caption{
Logarithm of mean number of photons $\lg(\bar{n})$ as a function of $\xi$ and $\lambda$. It can be seen that $\bar{n}$ increases with $\lambda$ and decreases with $\xi$.}
\label{fig2}
\end{figure}

A case of particular interest is the sideband transition, where the laser detuning $\Delta=\pm\omega_{\rm t}$, here $\omega_{\rm t}$ is the frequency of the trap. Because of AC-Stark shift and off-resonant transitions, the sideband Rabi frequency $\Omega_+$ has upper bound \cite{Steane2000}. Methods have been adopted to partially cancel the effect, and it seems feasible to have $\Omega_+<\omega_{\rm t}$ for special temporal and spectral arrangements of the laser field \cite{Haffner2008}. Since $\Omega_+=\frac{2\pi}{\lambda}\sqrt{\frac{\hbar}{2M\omega_{\rm t}}}\Omega$, where $M$ is the mass for a single ion, we have
\begin{equation}\label{ee2}
\Omega<\frac{\lambda}{2\pi}\sqrt{\frac{2M}{\hbar}}\omega_{\rm t}^{\frac{3}{2}}.
\end{equation}
From \cite{Wineland1998} and \cite{Steane1997}, it can be seen that
\begin{equation}
\eqalign{\Omega=-\frac{ea_0E}{4\hbar}=\frac{pE}{4\hbar},\cr
\omega_{\rm t}=\sqrt{\frac{e^2}{4\pi\epsilon_0Mz_{\rm s}^3}},}\label{e14}
\end{equation}
where $z_{\rm s}$ is the order of the separation between ions and is typically 10 to 100 $\mu$m.
Suppose $z_{\rm s}=\xi\lambda$, from (\ref{ee2}) and (\ref{e14}), we get
\begin{equation}  \label{ee5}
E<\frac{2\sqrt{2\hbar}}{p\pi}(\frac{e^2}{4\pi\epsilon_0})^{\frac{3}{4}}M^{-\frac{1}{4}}\xi^{-\frac{9}{4}}\lambda^{-\frac{5}{4}}.
\end{equation}
Substitute back to (\ref{b16}), we get
\begin{eqnarray}  \label{ee6}
\bar{n}&<\frac{3\epsilon_0^{\frac{1}{4}}}{32a_0^2\pi^{\frac{11}{4}}}\sqrt{\frac{\hbar}{e}}kM^{-\frac{1}{4}}\xi^{-\frac{9}{4}}\lambda^{\frac{7}{4}}\nonumber\\
&=6\times10^7kM^{-\frac{1}{4}}\xi^{-\frac{9}{4}}\lambda^{\frac{7}{4}}.
\end{eqnarray}
In the cases we consider, it is suitable to limit $k\leq2$,
9u~$\leq M\leq$~200u (u$=1.66057\times10^{-27}$~kg). For
$M=9$u, $k=2$, we get
\[\bar{n}=3.4\times10^{14}\xi^{-\frac{9}{4}}\lambda^{\frac{7}{4}}.
\]
We can see that a large $\lambda$ and a small $\xi$ result in a large $\bar{n}$. The curves of $\lg(\bar{n})$ is plotted in Figure~\ref{fig2} versus parameter $\xi$ from 2 to 100. When $\lambda=10^{-6}$ m and $\xi=2$, we get $\bar{n}=2.3\times10^3$.

There are also authors who have calculated $\bar{n}$ in a $k\pi$ pulse in another way \cite{Enk&Kimble2002}.To introduce this work, we first introduce the formalism developed by Blow {\it et al.} \cite{Blow1990}. This formalism is used to describe the continuous-mode coherent state with an arbitrary noncontinuous set of bases functions. Let $\phi_i(t)$ be a complete set of functions such that
\begin{eqnarray}
\int {\rm d} t \phi_i(t)\phi^*_j(t)=\delta_{ij},\nonumber\\
\Sigma_{i}\phi^*_i(t)\phi_i(t')=\delta(t-t').
\end{eqnarray}
The continuous-mode coherent state can be expressed as
\begin{equation} \label{e9}
|\alpha(\omega)\rangle=\exp\left(\int{\rm d}\omega[\alpha(\omega)a^\dag(\omega)-\alpha^*(\omega)a(\omega)]\right)|0\rangle,
\end{equation}
where $a^\dag(\omega)$ and $a(\omega)$ are continuous-mode creation and annihilation operators for each frequency $\omega$, $|0\rangle$ is the vacuum state and $\alpha(\omega)$ the continuous-mode coherent state amplitudes. In terms of this set, $|\alpha(\omega)\rangle$ can be expressed as a tensor product of coherent states $\bigotimes_i|\gamma_i\rangle$, where $\gamma_i$ is the eigenstate of a discrete annihilation operator, with the operator and eigenvalue functions of $\phi_i(t)$.

The authors of \cite{Enk&Kimble2002} consider the situation where a laser is used to drive Rabi oscillation of the atom, and take the laser as a continuous-mode coherent state. Using the formalism above, the interaction time $t$ of the field and ion can be expressed as $t=\int_{-\infty}^{\infty}{\rm d}\tau \phi(\tau)$. With an appropriate $\phi_i(t)$, they work out the interaction time for $k\pi$ pulse as
$
t=\frac{k\pi\hbar}{d}\sqrt{\frac{\epsilon_0cA}{2P}},$
where $d$ is the coupling constant of the atom and laser, and $P$ is the power of the laser. Thus, the mean number of photons in one $k\pi$ pulse is
$
\bar{n}\approx\frac{P}{\hbar\omega_{\rm L}}t=\frac{k\pi}{\omega_{\rm L} d}\sqrt{\frac{\epsilon_0cAP}{2}},
$
where $\omega_{\rm L}$ is the frequency of the representative single-mode coherent state. Thus, obviously, they take all the photons in area $A$ as effective photons when considering the interaction, but actually each photon in the beam does not have 100\% probability of interacting with the ion, thus the number of effective photons is much smaller.

\subsection{Accuracy of gate operation}

Suppose we have applied coherent pulses m times and reached
a state ${\rho}^{(m)}=\frac{1}{2}
({I}+{r}^{(m)}\cdot{\sigma})$. Let
$|\Psi\rangle=\alpha|0\rangle+\beta|1\rangle$ be the expected
state, the accuracy rate of gate operation realized through
Rabi oscillation is
\begin{eqnarray}
p_{\rm s}^{(m)}&=\langle\Psi|{\rho}^{(m)}|\Psi\rangle\nonumber\\
&=|\alpha|^2\rho_{11}^{(m)}+|\beta|^2\rho_{22}^{(m)}+\alpha^*\beta\rho_{12}^{(m)}+\alpha\beta^*\rho_{21}^{(m)}\nonumber\\
&=\frac{1}{2}\Big(1+r_{\rm z}^{(0)}r_{\rm z}^{(m)}+r_{\rm x}^{(0)}r_{\rm x}^{(m)}+r_{\rm y}^{(0)}r_{\rm y}^{(m)}\Big)\nonumber\\
&=\frac{1}{2}(1+{r}^{(0)}\cdot{r}^{(m)}),
\end{eqnarray}
for a mixed state, $|{r}^{(m)}|<1$, then $p_s<1$. The failure probability is
$
p_{\rm f}^{(m)}=1-p_{\rm s}^{(m)}.$
A detailed calculation results (see \ref{a8})
\begin{eqnarray*}
&p_{\rm f}^{(m)}=-\frac{1}{2}({r}^{(0)}\cdot{r}^{(m)}-1)\\
&=-\frac{1}{2}\Big\{(r^{(0)}_{\rm x})^2\big[(S_3+S_5)^m-1\big]+((r^{(0)}_{\rm y})^2+(r^{(0)}_{\rm z})^2)\big[|\lambda|^{\frac{m}{2}}
\cos(m\theta)-1\big]\\&+|\lambda|^{\frac{m}{2}}\sin(m\theta)
(\det{J})^{-\frac{1}{2}}\big[((r^{(0)}_{\rm y})^2-(r^{(0)}_{\rm z})^2)(a-d)+
r^{(0)}_{\rm y}r^{(0)}_{\rm z}(b+c)\big]\\&+B_1^{(m)}(r^{(0)}_{\rm y}c_{\rm y}+r^{(0)}_{\rm z}c_{\rm z})+B_2^{(m)}
\big[(r^{(0)}_{\rm y}c_{\rm y}-r^{(0)}_{\rm z}c_{\rm z})(a-d)+r^{(0)}_{\rm z}c_{\rm y}c+r^{(0)}_{\rm y}c_{\rm z}b\big]\Big\}.
\end{eqnarray*}

\begin{figure}[htbp]
\centering
\includegraphics[width=4.8in,height=3in]{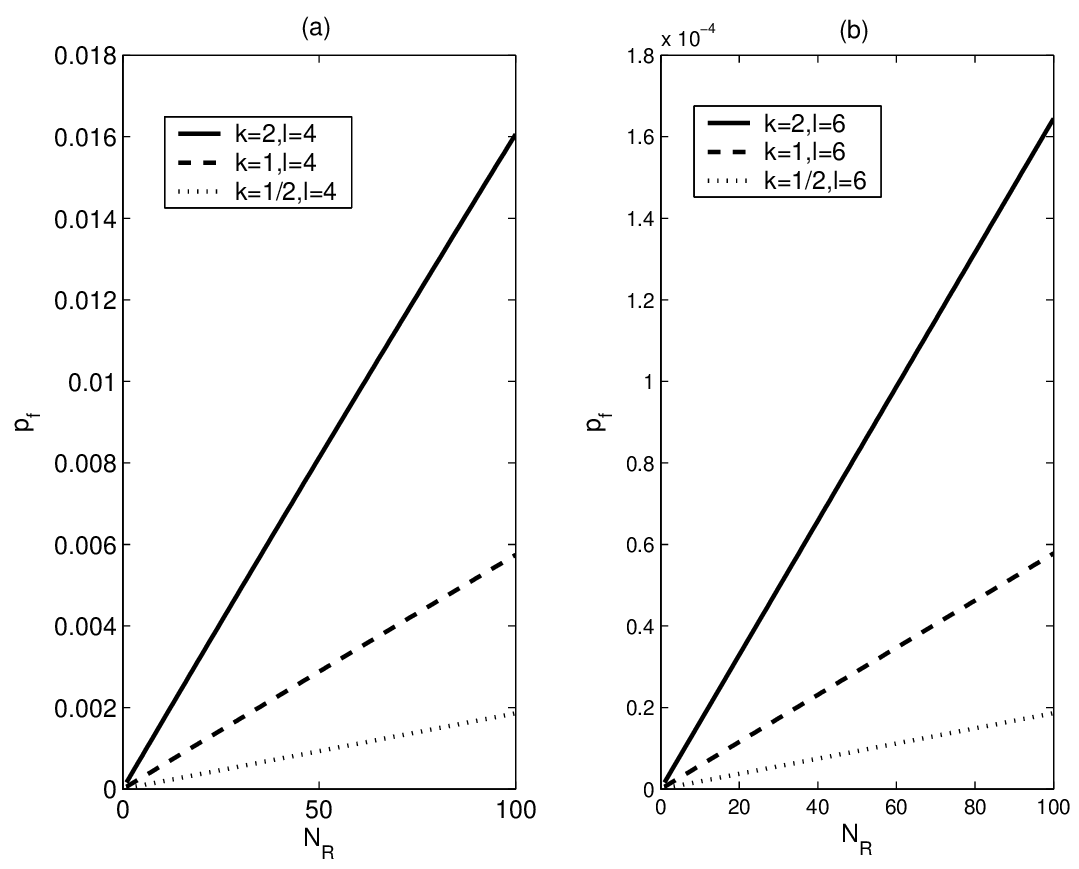}
\caption{
Failure probability $p_{\rm f}$ versus number of Rabi periods $N_{\rm R}$, $k\pi$ pulses are applied, and $\bar{n}$ is $10^4$ ($10^6$) in a (b). The failure probability $p_{\rm f}$ increases with the number of Rabi periods $N_{\rm R}$ and the value $k$, and is inversely proportional to $\bar{n}$.}
\label{fig1}
\end{figure}

We then average over all initial states of the ion, and get the
average failure probability. The failure probability for $k\pi$
pulses with different $\bar{n}$ are shown in Figure~\ref{fig1}.
It can be seen that the failure probability increases with the
number of Rabi periods $N_{\rm R}$ and the value $k$, and is
inversely proportional to $\bar{n}$.

\section{Discussions}\label{a4}

\subsection{The permitted depth of quantum logical operation}

The failure probability we have calculated for the $\pi$ sideband
transition is $10^{-2}$ after approximately $10^2$
operations when $\bar{n}=10^4$, and after one
operation the failure probability is $10^{-4}$ under the same conditions. Gea-Banacloche has pointed out that\cite{Julio2002} for one Hadamard
transformation driven by a coherent field, the failure
probability from quantization of laser field is about
$0.22/\bar{n}$. However, his quantization is to add quantum fluctuations to
classical treatment of the laser field. Whether the result is still valid after many operations is not yet clear. In addition, compared with the transformation driven by $\pi$ pulses, the Hadamard transformation may have a smaller failure probability.

For controlled-NOT (CNOT)
gates, there are five steps in the Cirac--Zoller scheme, and two
steps are realized via Rabi oscillations driven by $\pi$
pulses. Generally speaking, the failure probability after five
steps is not less than that after one $\pi$ pulse. Then the
failure probability after repeated $\pi$ pulses is a lower
bound of the failure probability after repeated Cirac-Zoller's
CNOT gate. Thus the lower bound of the failure probability is
$10^{-2}$ after approximately $10^2$ CNOT operations when
$\bar{n}=10^4$.

The threshold theorem in QC declares that an
arbitrarily long computation can be performed reliably
if the failure probability of each quantum gate is less than a
critical value. Knill has used numerical calculations and obtained
a failure probability threshold of the order $10^{-2}$ \cite{Knill2005} based on a
fault-tolerant structure suggested by himself. P. Aliferis {\it
et al.} has reached a threshold of $10^{-3}$ with provable
constructions \cite{Aliferis2008}.

A parameter called permitted depth of logical operation
describing the property of a physical realization scheme of QC
has been given \cite{Yang2007}: considering that different
number state components of the driving field lead to different
oscillation amplitudes, which become gradually uncorrelated, we
can see that the failure probability of quantum logic gates has a
theoretical limitation. Combining this limitation given by the
quantum nature of the field with the threshold theorem in FTQC,
we can obtain the permitted depth of logical operation. This parameter
limits the number of operations on any physical qubit in one
error-correction period. Then the permitted depth of logical
operation here is less than $10^2$.


\subsection{Others' proposals which may have different results}

For a Rabi oscillation driven by microwaves, the failure probability may be much smaller because of a large mean number of photons, but it becomes difficult to individually address each of the ions. Although an additional magnetic field gradient applied to an electrodynamic trap may individually shift ionic qubit resonances \cite{Mintert2001}, thus making them distinguishable in frequency space, whether it can improve the permitted depth of logical operation needs further investigation.

There exists a two-qubit gate scheme totally different from the Cirac--Zoller gate, namely the scheme implemented by the NIST group \cite{Leibfried2003}. In this scheme off-resonant excitations of the stronger carrier transition are absent, and this allows a greater gate speed and thus a higher laser intensity. Besides, additional Stark shifts can be efficiently suppressed by choosing almost perpendicular and linear polarizations for the laser beams \cite{Wineland2003}. Hence, studies on this type of gate may lead to different results.

\section{Conclusions}\label{a5}

Firstly, we have investigated Rabi
oscillation of a two-level system driven by a pulse train. We developed an algorithm to solve the infinite summation, with a higher precision than has ever been reached. We have found that in this kind of Rabi oscillation there is a dual-pulse
structure in every period. The envelope of
population inversion collapses exponentially, unlike a
Gaussian function collapse envelope driven by a cw field.
Besides, there is no revival phenomenon, but a small nonzero
amplitude exists (the amplitude of each dual-pulse structure's crest approaches to a stable value).

Secondly, we have considered the application to gate operation in ion trap QC. We gave a lower bound of failure
probability. Our result is: when the wavelength of the
driving field is of the order $10^{-6}$ m, the mean number of
photons cannot be greater than $10^4$. Then, after about $10^2$ CNOT
gates in the Cirac--Zoller scheme, the lower bound of failure
probability is of the order $10^{-2}$.



\ack We thank Zheng-Wei Zhou, Yong-Sheng Zhang, Li You, Biao
Wu, Duan-Lu Zhou, and Chong Xiang for useful discussions. This
work was supported by the National Natural Science Foundation
of China under Grant No. 61173157.

\appendix

\section{PROOF OF PRECISION OF THE ALGORITHM IN SEC. \ref{b1}} \label{a6}
{\bf Proof of Lemma 1:}
For
$\bar{n}-\alpha\sqrt{\bar{n}}<n<\bar{n}+\alpha\sqrt{\bar{n}}$,
i.e.
$-\frac{\alpha}{\sqrt{\bar{n}}}<x<\frac{\alpha}{\sqrt{\bar{n}}}$,
after expanding $f_{i1} (x,\bar{n},k)$ at $x=0$, we get the result
$f_{i2} (x,\bar{n},k)$ satisfying $f_{i2} (x,\bar{n},k)= f_{i0}
(n,\bar{n},k) + o (x^p)$. It can be seen that $f_{i2}
(x,\bar{n},k)=f_{i3} (n, \bar{n},k)$, thus we have $f_{i0}
(n,\bar{n},k)=f_3 (n, \bar{n},k) + o (x^p)$, then
Eq.~(\ref{e23}) is proved ~$\Box$

{\bf Proof of Lemma 2:}
For every given $n$ satisfying $n<k+1<\bar{n}$, we
have $\frac{\bar{n}^j}{n!}<\frac{\bar{n}^{k+1}}{(k+1)!}$, thus
\begin{eqnarray}
\sum_{n=0}^ke^{-\bar{n}}\frac{\bar{n}^j}{n!}<\sum_{n=0}^ke^{-\bar{n}}\frac{\bar{n}^{k+1}}{(k+1)!}
=e^{-\bar{n}}\frac{\bar{n}^{k+1}}{k!}.
\end{eqnarray}
From Stirling's formula $k!=\sqrt{2\pi
k}(\frac{k}{e})^ke^{\frac{\theta}{12k}}, 0<\theta<1,$ we have
\begin{equation}  \label{e24}
e^{-\bar{n}}\frac{\bar{n}^{k+1}}{k!}<(\frac{e}{k})^ke^{-\bar{n}}\bar{n}^{k+1}=e^{k-\bar{n}}\bar{n}(\frac{k}{\bar{n}})^{-k}.
\end{equation}
Substitute $k$ in formula (\ref{e24}) with
$\bar{n}-\alpha\sqrt{\bar{n}}$, we have
\begin{eqnarray*}
e^{k-\bar{n}}\bar{n}(\frac{k}{\bar{n}})^{-k}&=\bar{n}e^{-\alpha\sqrt{\bar{n}}}(1-\frac{\alpha}{\sqrt{\bar{n}}})^{-\frac{\sqrt{\bar{n}}}{\alpha}(\alpha\sqrt{\bar{n}}-\alpha^2)}\nonumber\\
&=\bar{n}e^{-\alpha\sqrt{\bar{n}}}e^{\alpha\sqrt{\bar{n}}-\alpha^2}=\bar{n}e^{-\alpha^2}.
\end{eqnarray*}
When $\alpha>\sqrt{(l+1)\ln\bar{n}}$, we have
$\bar{n}e^{-\alpha^2}<\frac{1}{\bar{n}^l}$,
inequality~(\ref{e4}) is proved.~$\Box$

{\bf Proof of Lemma 3:} It can be seen that
\begin{eqnarray*}
\sum_{n=k'}^{\infty}e^{-\bar{n}}\frac{\bar{n}^n}{n!}&=e^{-\bar{n}}\frac{\bar{n}^{k'}}{k'!}\sum_{n=k'}^{\infty}\frac{\bar{n}^{n-k'}k'!}{n!}\nonumber\\
&<e^{-\bar{n}}\frac{\bar{n}^{k'}}{k'!}\sum_{n=k'}^{\infty} (\frac{\bar{n}}{k'+1})^{n-k'}
=e^{-\bar{n}}\frac{\bar{n}^{k'}}{k'!}\frac{k'+1}{k'+1-\bar{n}},
\end{eqnarray*}
when $k'>\bar{n}+\frac{1}{\bar{n}}$, i.e,
$\frac{k'+1}{k'+1-\bar{n}}<k'$, we have
\[
e^{-\bar{n}}\frac{\bar{n}^{k'+1}}{k'!}<(\frac{e}{k'})^{k'}e^{-\bar{n}}\bar{n}^{k'+1}<\bar{n}e^{-\bar{n}}\frac{\bar{n}^{k'-1}}{ (k'-1)!}.
\]
with Stirling's formula we get
\[
(\frac{e}{k'})^{k'}e^{-\bar{n}}\bar{n}^{k'+1}<\bar{n}e^{-\bar{n}}\frac{\bar{n}^{k'-1}}{ (k'-1)!}<\bar{n}e^{-\bar{n}}\bar{n}^{k'-1}=\bar{n}e^{-\bar{n}}\bar{n}^{k'-1}.
\]
Let $\lambda=\frac{\bar{n}}{k'-1}<1, \eta=\bar{n}^{\frac{l+1}{\bar{n}}}$, we then have
\begin{eqnarray*}
\bar{n}e^{-\bar{n}}\bar{n}^{k'-1}=\bar{n}e^{-\bar{n}}\bar{n}^{k'-1}<\frac{1}{\bar{n}^l}&\Leftrightarrow
(\frac{\bar{n}e}{k'-1})^{k'-1}<\frac{e^{\bar{n}}}{\bar{n}^{l+1}}\Leftrightarrow(\frac{e}{\eta})^{\lambda}-e\lambda>0\nonumber\\
&\Leftrightarrow\lambda(1-\ln\eta)>1+\ln\lambda.
\end{eqnarray*}
Let $\lambda=1-\Delta$, with $0<\Delta<1$, from
$\ln(1+x)<x-\frac{1}{2}x^2$ $(x<0)$, we get
$\ln\lambda<-\Delta-\frac{1}{2}\Delta^2$, then a sufficient
condition of $(\frac{e}{\eta})^{\lambda}-e\lambda>0$ is:
\[
(1-\Delta)(1-\ln\eta)>1-\Delta-\frac{1}{2}\Delta^2,
\]
which results in
$\Delta>\Delta_0$, here $\Delta_0\equiv-\ln\eta+\sqrt{\left(\ln\eta\right)^2+2\ln\eta}$.
Let $\frac{\bar{n}}{k_0-1}=\bar{n}+\alpha_0\sqrt{\bar{n}}$, we
get
\begin{eqnarray*}
\alpha_0&=\frac{1}{\sqrt{\bar{n}}}\left[\left(\frac{1}{1-\Delta_0}-1\right)\bar{n}+1\right]\\
&=\frac{1}{\sqrt{\bar{n}}}+\left(\frac{l+1}{\bar{n}}\ln\bar{n}+\sqrt{\left(\frac{l+1}{\bar{n}}
\ln\bar{n}\right)^2+2\frac{l+1}{\bar{n}}\ln\bar{n}}\right)\sqrt{\bar{n}}
\end{eqnarray*}
by using
$\frac{1}{1-\Delta_0}=1+\ln\eta+\sqrt{\left(\ln\eta\right)^2+2\ln\eta}$
and $\eta=\left(\bar{n}\right)^\frac{l+1}{\bar{n}}$. Because
$\Delta>\Delta_0\Leftrightarrow\alpha>\alpha_0$, we get a
sufficient condition of Lemma 3:
\begin{eqnarray*}
\alpha&>\frac{1}{\sqrt{\bar{n}}}+\left(\frac{l+1}{\bar{n}}\ln\bar{n}+\sqrt{\left(\frac{l+1}{\bar{n}}
\ln\bar{n}\right)^2+2\frac{l+1}{\bar{n}}\ln\bar{n}}\right)\sqrt{\bar{n}}\nonumber\\
&=\frac{1}{\sqrt{\bar{n}}}+\frac{(l+1)\ln\bar{n}}{\sqrt{\bar{n}}}+\sqrt{\frac{(l+1)^2(\ln\bar{n})^2}{\bar{n}}+2(l+1)\ln\bar{n}}.
\end{eqnarray*}
then Lemma 3 follows.~$\Box$

{\bf Proof of Theorem 1:} From {\bf Lemma 1,2} and {\bf 3} we get: for every
given $l<<\bar{n}$, if $\alpha$ satisfies

\begin{eqnarray}
&~~~~~\frac{1}{\sqrt{\bar{n}}}
+\frac{(l+1)\ln\bar{n}}{\sqrt{\bar{n}}}+\\ \nonumber
&+\sqrt{\frac{(l+1)^2(\ln\bar{n})^2}{\bar{n}}+2(l+1)\ln\bar{n}}<\alpha<<\sqrt{\bar{n}},
\end{eqnarray}

(\ref{e23}), (\ref{e4}) and (\ref{e8}) hold. Then
\begin{eqnarray*}
S_i&=\sum_{n=0}^{\infty}\frac{e^{-\bar{n}}\bar{n}^n}{n!}f_{i0}(n,\bar{n},k)\nonumber\\
&=\sum_{n=0}^{\bar{n}-\alpha\sqrt{\bar{n}}}\frac{e^{-\bar{n}}\bar{n}^n}{n!}(f_{i0}(n,\bar{n},k)-f_{i3}(n,\bar{n},k))\nonumber\\
&+\sum_{n=\bar{n}+\alpha\sqrt{\bar{n}}}^{\infty}\frac{e^{-\bar{n}}\bar{n}^n}{n!}(f_{i0}(n,\bar{n},k)-f_{i3}(n, \bar{n},k))
+\sum_{n=0}^{\infty}\frac{e^{-\bar{n}}\bar{n}^n}{n!}f_{i3}(n, \bar{n},k)\nonumber\\
&+\sum_{n=\bar{n}-\alpha\sqrt{\bar{n}}}^{\bar{n}+\alpha\sqrt{\bar{n}}}\frac{e^{-\bar{n}}\bar{n}^n}{n!}(f_{i0}(n, \bar{n},k)-f_{i3}(n, \bar{n},k))\nonumber\\
&=\sum_{n=0}^{\infty}\frac{e^{-\bar{n}}\bar{n}^n}{n!}f_{i3}(n, \bar{n},k)+o\Big(\frac{1}{\bar{n}^l}\Big)+o\Big(\frac{\alpha^{p+1}}{(\sqrt{\bar{n}})^{p-1}}\Big).
\end{eqnarray*}
Let $l'$ satisfy
$\frac{1}{\bar{n}^{l'}}>\frac{\alpha^{p+1}}{(\sqrt{\bar{n}})^{p-1}}$,
we get
\[
l'>\frac{(p+1)(\frac{1}{2}\ln\bar{n}-\ln\alpha)-\ln\bar{n}}{\ln\bar{n}}.
\]
If $l<l'$, then we have $\frac{1}{\bar{n}^l}>\frac{1}{\bar{n}^{l'}}>\frac{\alpha^{p+1}}{(\sqrt{\bar{n}})^{p-1}}$. From Eq.~(\ref{e10}) we get $\alpha>\sqrt{2(l+1)\ln\bar{n}}$, then
\[p\geq\left\lceil \frac{2\ln\Big(\sqrt{2}\bar{n}^{l+\frac{1}{2}}\sqrt{(l+1)\ln\bar{n}}\Big)}{\ln\bar{n}-\ln[2(l+1)\ln\bar{n}]}
\right \rceil,\] then
$S_i=\sum_{n=0}^{\infty}\frac{e^{-\bar{n}}\bar{n}^n}{n!}f_{i3}(n,
\bar{n},k)+o\Big(\frac{1}{\bar{n}^l}\Big).$ Since we can get
exact result of
$\sum_{n=0}^{\infty}\frac{e^{-\bar{n}}\bar{n}^n}{n!}f_{i3}(n,
\bar{n},k)$, we get $S_i$ with precision
$o\Big(\frac{1}{\bar{n}^l}\Big)$.~$\Box$

\section{CALCULATION OF ${M_1}^m$} \label{a7}

Let
\begin{equation} \label{e20}
{M}=\left[\begin{array}{ccc}
S_3+S_5 & 0\\
{O} &{M}_1 \\
\end{array}\right],
\end{equation}
where \[{M}_1=\left[\begin{array}{ccc}
S_5-S_3&- (S_1+S_7) \\
S_2&S_4+S_6-1\\
\end{array}\right]\stackrel{\triangle}{=}\left[\begin{array}{ccc}
~a~&b~ \\
~c~&d~\\
\end{array}\right],\]
then
\[{M}^m=\left[\begin{array}{ccc}
(S_3+S_5)^m & 0\\
{O} &{M}_1^m \\
\end{array}\right].\]

\begin{figure}
 \centering
  \includegraphics[width=3in]{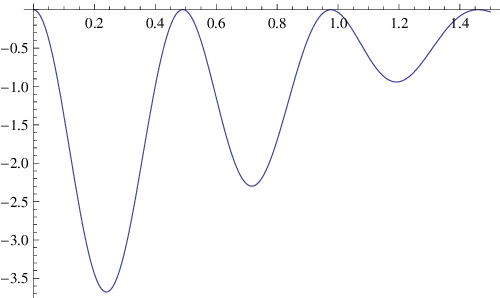}\\
  \caption{
  $\Delta(\tau)=(a-d)^2+4bc<0$ versus $\tau=gt$, where $t$ is the pulse width. Different $\Delta(\tau)$ results in different behavior of Rabi oscillation driven by pulse train. For the cases we consider, $\tau<1$, we can see $\Delta(\tau)<0$.}
  \label{fig3}
\end{figure}

Let $(1, x_{21})^T$ and $(1, x_{22})^T$ be the eigenvectors of ${M}_1$ with corresponding eigenvalues $\lambda_1$ and $\lambda_2$, then
\begin{eqnarray*}
&x_{21}=\frac{1}{2b}[d-a+\sqrt{(a-d)^2+4bc}],\\
&x_{22}=\frac{1}{2b}[d-a-\sqrt{(a-d)^2+4bc}],\\
&\lambda_1=\frac{1}{2}[a+d+\sqrt{(a-d)^2+4bc}],\\
&\lambda_2=\frac{1}{2}[a+d-\sqrt{(a-d)^2+4bc}],
\end{eqnarray*}
we have plot $\Delta(\tau)=(a-d)^2+4bc$ versus $\tau=gt$ in Figure~\ref{fig3}.
$\Delta(\tau)$ is below zero for the cases we are interested in ($\tau<1$).

Let
\[
{T}_1=\left[\begin{array}{lll}
1&1 \\
x_{21}&x_{22}\\
\end{array}\right],
\]
thus
\[
{M}_1={T}_1\left[\begin{array}{lll}
\lambda_1&0 \\
0&\lambda_2\\
\end{array}\right]{T}_1^{-1},
\]
then,
\[
\fl
{M}_1^m={T_1}\left[\begin{array}{lll}
\lambda_1^m&0 \\
0&\lambda_2^m\\
\end{array}\right]{T_1}^{-1}=\frac{1}{x_{21}-x_{22}}\left[\begin{array}{lll}
-x_{22}\lambda_1^m+x_{21}\lambda_2^m&\lambda_1^m-\lambda_2^m \\
x_{21}x_{22}(-\lambda_1^m+\lambda_2^m)&x_{21}\lambda_1^m-x_{22}\lambda_2^m\\
\end{array}\right].
\]

Denote $\lambda_1^m\pm\lambda_2^m=\Lambda_{\pm}^{(m)}, d-a=K, \sqrt{(a-d)^2+4bc}={\rmi}Q$, we have
\[{M}_1^m=\frac{\Lambda_{+}^{(m)}}{2}{I}+\frac{\Lambda_{-}^{(m)}}{2\rmi Q}\left[\begin{array}{lll}
-K&2b\\
2c&K\\
\end{array}\right].
\]
It can be seen that $|\lambda_1|=|\lambda_2|$, let $\lambda_1=|\lambda|e^{\rmi\theta}, \lambda_2=|\lambda|e^{-\rmi\theta}$, using $|\lambda|^2=\lambda_1\lambda_2=ad-bc$, we get
\begin{eqnarray*}
&\Lambda_{+}^{(m)}=2(ad-bc)^{m/2}\cos(m\theta),\\
&\Lambda_{-}^{(m)}=2\rmi(ad-bc)^{m/2}\sin(m\theta),
\end{eqnarray*}
where $\theta$ satisfies $\sin\theta=\sqrt{\frac{2ad-4bc-a^2-d^2}{4(ad-bc)}}$,
then
\begin{equation} \label{e21}
{M}_1^m=|\lambda|^m\Big[\cos(m\theta){I}+\sin(m\theta)
\frac{{J}}{\sqrt{\det{J}}}\Big],
\end{equation}
where \[{J}=\left[\begin{array}{ccc}
a-d&2b\\
2c&d-a\\
\end{array}\right].\]

\section{CALCULATION OF ${r}^{(0)}\cdot{r}^{(m)}$} \label{a8}

It can be seen from (\ref{e19}), (\ref{e22}), (\ref{e20}), (\ref{e21}) that
{\footnotesize
\begin{eqnarray} \label{e16}
\fl
{r}^{(0)}\cdot{r}^{(m)}=({r}^{(0)})^T
{M}^m{r}^{(0)}+({r}^{(0)})^T\left[\begin{array}{lll}
\sum _{k=0}^{m-1 }(S_3+S_5)^k & 0\\
{O} &\sum _{k=0}^{m-1 }{M}_1^k \\
\end{array}\right]{c}\nonumber\\
=({r}^{(0)})^T
\left[\begin{array}{lll}
(S_3+S_5)^m & 0\\
{O} &{M}_1^m \\
\end{array}\right]{r}^{(0)}+({r}^{(0)})^T\left[\begin{array}{lll}
\frac{1-(S_3+S_5)^m}{1-(S_3+S_5)} & 0\\
{O} &B_1^{(m)}{I}+B_2^{(m)}{J} \\
\end{array}\right]{c}\nonumber\\
=(r^{(0)}_{\rm x})^2(S_3+S_5)^m+B_1^{(m)}(r^{(0)}_{\rm y}c_{\rm y}+r^{(0)}_{\rm z}c_{\rm z})+
((r^{(0)}_{\rm y})^2+(r^{(0)}_{\rm z})^2)|\lambda|^{\frac{m}{2}}\cos(m\theta)\nonumber\\
+|\lambda|^{\frac{m}{2}}\frac{\sin(m\theta)}{\sqrt{\det{J}}}
\left[\begin{array}{lll}
r^{(0)}_{\rm y} & r^{(0)}_{\rm z}\\
\end{array}\right]{J}\left[\begin{array}{lll}
r^{(0)}_{\rm y}\\
r^{(0)}_{\rm z} \\
\end{array}\right]+B_2^{(m)}\left[\begin{array}{lll}
(r^{(0)}_{\rm y} & r^{(0)}_{\rm z}\\
\end{array}\right]{J}\left[\begin{array}{lll}
c^{(0)}_{\rm y}\\
c^{(0)}_{\rm z} \\
\end{array}\right].
\end{eqnarray}
}
Using \[
\left[\begin{array}{lll}
r^{(0)}_{\rm y} & r^{(0)}_{\rm z}\\
\end{array}\right]{J}\left[\begin{array}{lll}
c^{(0)}_{\rm y}\\
c^{(0)}_{\rm z} \\
\end{array}\right]=(r^{(0)}_{\rm y}c_{\rm y}-r^{(0)}_{\rm z}c_{\rm z})(a-d)+r^{(0)}_{\rm z}c_{\rm y}c+r^{(0)}_{\rm y}c_{\rm z}b,
\]
and
\[
\left[\begin{array}{lll}
r^{(0)}_{\rm y} & r^{(0)}_{\rm z}\\
\end{array}\right]{J}\left[\begin{array}{lll}
r^{(0)}_{\rm y}\\
r^{(0)}_{\rm z} \\
\end{array}\right]=((r^{(0)}_{\rm y})^2-(r^{(0)}_{\rm z})^2)(a-d)+r^{(0)}_{\rm y}r^{(0)}_{\rm z}(b+c),
\]
we get
{\footnotesize
\begin{eqnarray*}
\fl
{r}^{(0)}\cdot{r}^{(m)}=
(r^{(0)}_{\rm x})^2\big[(S_3+S_5)^m-1\big]+((r^{(0)}_{\rm y})^2+(r^{(0)}_{\rm z})^2)\big[|\lambda|^{\frac{m}{2}}
\cos(m\theta)-1\big]\\+|\lambda|^{\frac{m}{2}}\sin(m\theta)
(\det{J})^{-\frac{1}{2}}\big[((r^{(0)}_{\rm y})^2-(r^{(0)}_{\rm z})^2)(a-d)+
r^{(0)}_{\rm y}r^{(0)}_{\rm z}(b+c)\big]\\+B_1^{(m)}(r^{(0)}_{\rm y}c_{\rm y}+r^{(0)}_{\rm z}c_{\rm z})+B_2^{(m)}
\big[(r^{(0)}_{\rm y}c_{\rm y}-r^{(0)}_{\rm z}c_{\rm z})(a-d)+r^{(0)}_{\rm z}c_{\rm y}c+r^{(0)}_{\rm y}c_{\rm z}b\big]+1.
\end{eqnarray*}
}

\end{document}